\documentclass[12pt, article]{aastex}
\usepackage{natbib}
\begin{document}
\title{Comet Encounters and Carbon 14}
\author{David Eichler}
\affil{Physics Department, Ben-Gurion University, Be'er-Sheva 84105, Israel\\
        E-mail: \email{eichler.david@gmail.com}}
\author{David Mordecai}

\begin{abstract} The $^{14}$C production of  shock-accelerated particles is calculated in terms of the total energy released in energetic particles. The recently reported 1.2\% jump in the $^{14}$C content of the atmosphere in the year C.E. 775, it is found, would require $\gtrsim 10^{34}$ erg in energetic particles, less than first estimates but far more than any known solar flare on record. It is noted that the superflare from a large comet (comparable to C/Hale-Bopp) colliding with the sun could produce shock-accelerated GeV cosmic rays in the solar corona and/or solar wind, and possibly account for the CE 775 event. Several additional predictions of cometary encounters with the sun and other stars may be observable in the future.
\end{abstract}


 \section{Introduction}
 Comets, asteroids and giant solar flares each pose dangers. If a comet, or its coma or tail were to sufficiently rattle the Earth's magnetosphere, the electromagnetic disturbance so induced could threaten modern civilization, which depends on functioning microelectronics. A sufficiently  large sun-grazing comet (or asteroid) $R\gtrsim 10^{7}$ cm,  would
 have a mass of $10^{21.5}$ g and
 contain over $10^{36}$ ergs, more than two orders of magnitude more kinetic energy near the sun than a reasonable estimate for the energy in the Carrington solar flare of 1859 (which damaged early telegraph lines), and its energy release near the sun's surface could traumatize the Earth with UV exposure and electromagnetic disturbance.
 A mid-size to large comet ($R\gtrsim 3$ km) impacting Earth could deposit $\gtrsim 10^{30}$ ergs into the ocean, enough to supersaturate the Earth's atmosphere with water vapor, leading to something resembling legendary floods, and could dramatically heat at least parts of the atmosphere.


 Sun-grazing comets are  a fact of life (e.g. Schrijver, et. al 2012, Sekanina and Chodas, 2012).  The possibility clearly exists that some of them could be, at some stage, quite large ($M \gg 10^{19}$ g) and collide with the sun, causing an explosive release of more than   $10^{34}$ erg in energy (Brown et al., 2011). The lack of any known, reliable record of a major cataclysm  associated with  such past events  could be interpreted - depending  on the size and kinetic energy of the comet - as  evidence that they were less conspicuous to a pre-electronic civilization than one might suppose, or that superstitions interpreting comets as bad omens had some historical basis.  In any case, they could be far more disruptive to post-20th century civilization, and their event rate, even if small, is worthy of study. 

Here we consider whether a) a giant solar flare, or b) the close approach of a large comet to the sun  could have occurred in the year 775, when the levels of $^{14}$C rose by 1.2 percent within a year or so (Miyake, et al. 2012).
 This rise would require 10 years worth of normal cosmic ray exposure within one year. While it could have been due to an extremely large but otherwise normal solar flare, its statistical deviation from other  $^{14}$C rises on record motivates us to consider  whether a different type of event from normal solar flares could produce the $^{14}$C enhancement. If it was indeed due to a giant comet closely encountering the sun, it would be evidence that such an event is survivable, and quantitative estimates of the energy  it released are desirable. By the same token, it is worth considering whether a less close approach could have produced the enhancement by tapping the energy in the solar wind to produce a temporary rise in energetic particles.





\section{Particle Acceleration and $^{14}$C Production}


   The  energetic particle spectrum F(p)dlnp at momentum p for  shock-accelerated ions at the shock is given as
   \begin{eqnarray} F(p) &\propto &   \exp  \int_{min}^{p}{3\over 2(r^{-1}-1)}\bigg[\left(-1+[1+{2\pi^2D_{\parallel}(p')D_{\perp}(p')\over{R_s^2u_s^2}}]^{1/2}\right) \nonumber \\     &&+ \left(r^{-1} +[r^{-2}+{2\pi^2D_{\parallel}(p')D_{\perp}(p')\over{R_s^2u_s^2 }}]^{1/2}\right)\bigg]dlnp  
   \label{spectrum}
  \end{eqnarray}
\label{spectrum}
   (Eichler, 1981a).  
    Here r is the compression ratio of the shock, $\left[D_{\parallel}D_{\perp}\right]^{1/2} =  pvc/2^{1/2}\pi\eta ZeB$ is the geometric mean of the parallel and perpendicular diffusion coefficients of ions of momentum p, velocity v, and charge Ze in the magnetic field B,  $R_s$ is the radius of the shock, and $u_s$ is its velocity relative to the pre-shock fluid. The spectrum cuts off exponentially at energy
   $E\ge E_o\equiv
   \eta ZeBR_su_s/c$, where     the dimensionless coefficient $\eta $ is about 1/3, based on observations of the energetic particles at the Earth's bow shock (Ellison and Mobius, 1987, Chang et al, 2001)
   at which $R_s \simeq 6 \cdot 10^9$ cm and $E_o \simeq 36 Z$ KeV.  For $pv \ll E_o$, the integral spectral index s is -1 for a strong shock of compression ratio 4 in the test particle approximation.

   The spectrum of {\it escaping} particles, in the approximation of steady state, is proportional to their rate of production at the shock, which is proportional to $u_s(1-1/r)(dF/dlnp)/3$. This expression includes particle escape by both convection and diffusion to a free streaming boundary. We assume that the  particles mostly responsible for the $^{14}$C, i.e. the most energetic,  freely stream from a sunward acceleration site towards Earth and precipitate onto its polar caps. Particles trapped in the expanding flow are those at low energy, and their adiabatic losses can be recycled into the acceleration of the expanding blast wave.

   The $^{14}$C production rate  per unit particle energy $ \int E {dF\over dlnp}dlnp$ in the Earth's atmosphere is independent of the normalization of F and is given by
 \begin{equation}
Q  =  \int{Y(E)\over \pi}{\Omega(E)\over 2\pi} {dF(p)\over dln p}d ln p/\int E {dF\over dlnp}dlnp
 \label{c14rate}
 \end{equation}
 where E is the kinetic energy, hereafter expressed in units of $m_pc^2$,  $Y(E)/\pi$, the neutron yield per primary cosmic ray of energy E, can be approximated as $4 E^{2.35}/(E^2 +E^{0.35}) $   (Kovaltsov et al., 2012 ),  
   $\Omega(E)/(2\pi)  \simeq 0.13 E^{1/4}$ is the solid angle   of the magnetic polar cap whose geomagnetic cutoff is E. (Here we have assumed that the angular distribution of the energetic particles is energy independent.)
Using equations (\ref{spectrum}) and (\ref{c14rate}), we numerically calculate the total $^{14}$C yield per unit energy as a function of the parameter $E_o$. For reference, note that the most energy-efficient ion energy for making $^{14}$C is at the maximum of $\Omega(E)Y(E)/E$,  and that the yield is $\sim 0.3$ $^{14}$C atoms per $m_pc^2$ of kinetic energy. The results for an actual shock-accelerated spectrum over a wide range of $E_o$ are plotted in figure 1, showing that over a wide range of $E_o$, the  $^{14}$C production  efficiency $\epsilon$ is within a factor of several of the maximum.

\begin{figure}
\includegraphics[scale=0.5]{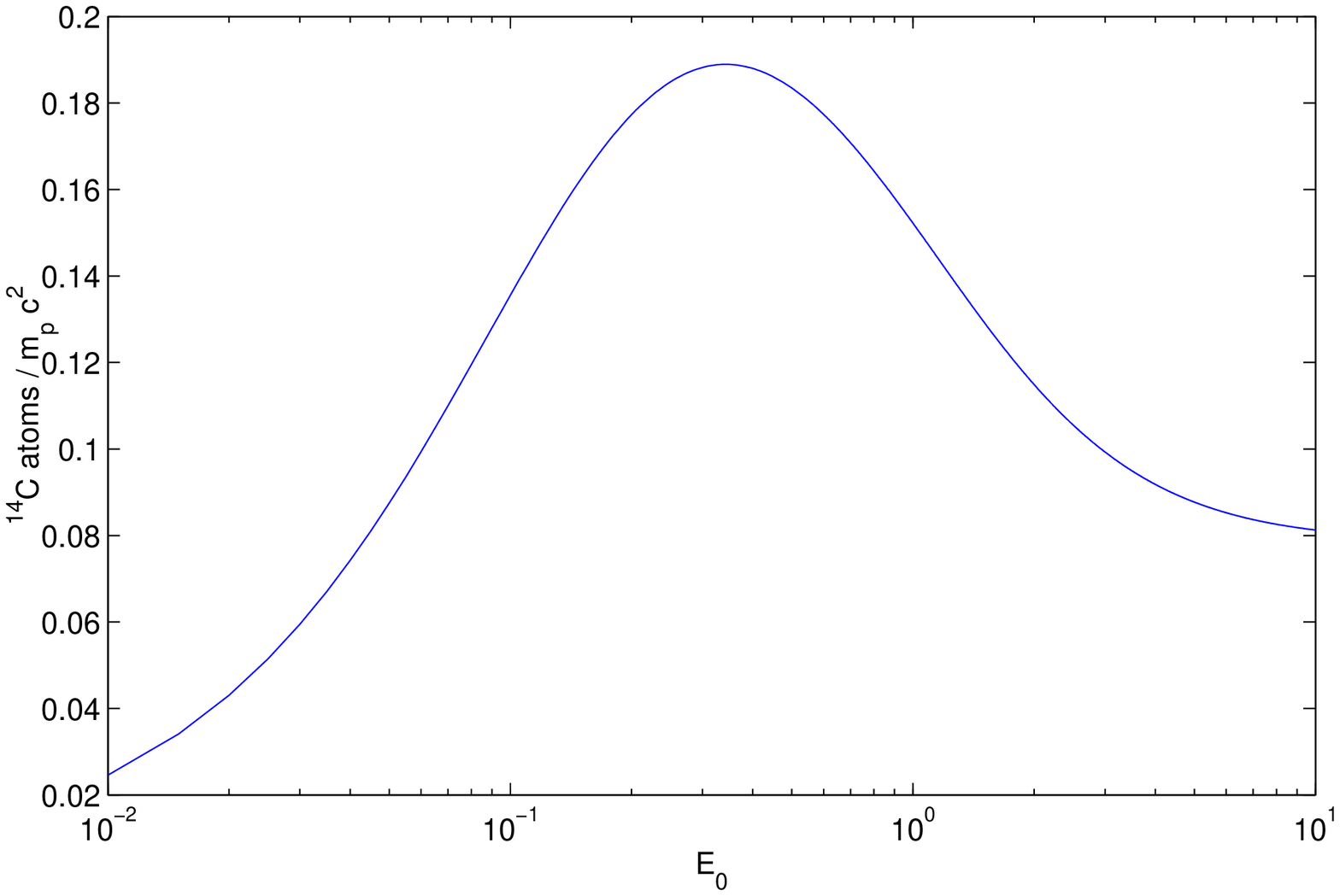}
\caption{total $^{14}$C yield per total kinetic energy as a function of the spectral cutoff parameter $E_0$ }
\end{figure}

An episode in which $6 \cdot 10^8$ $^{14}$C atoms/cm$^2$ are produced at Earth then requires a (kinetic) energy fluence at Earth of $\sim 6\epsilon^{-1}\cdot 10^8 m_p c^2$/cm$^2$. A giant particle acceleration event near the sun that spewed out energetic particles over a solid angle of $\sim 2\pi$  would thus have needed to produce $8.4 \cdot 10^{35}\epsilon^{-1} m_pc^2 \simeq 1.3\cdot 10^{33} \epsilon^{-1}$ ergs in energetic particles. The total energy of the event, of course, could be more, though several observations in the heliosphere (e.g. Eichler, 1981b, Ellison \& Mobius, 1987, Decker et. al, 2008) support the theoretical assertion (Eichler, 1979, Eichler, 1985, Ellison \& Eichler, 1985) that energetic particles can contain a significant fraction of the total energy in a collisionless blast.
   The value of the cutoff energy  $E_o=eBRu_s/3c$ obtained by using solar parameters  -
       $ B\gtrsim 0.3$ G, $R \sim 6\cdot 10^{10}$ cm, $u_s/c \gtrsim 10^{-3}$ - is comfortably above 1 GeV, so, even if the lateral extent of the shock is somewhat less than the above value of R, we may reasonably assume a value of $\epsilon$ in the range 0.1 to 0.2. A superflare yielding  $\sim 10^{34}$ ergs in energetic particles could  thus account for  the $^{14}$C enhancement of 775.  For $E_o\gtrsim 0.3m_pc^2$, this energy requirement is less than the estimate in Miyake et al. (2012), presumably because the particle spectrum predicted here is more efficient for $^{14}$C production than the hardest recorded solar flare spectra, which was adopted by those authors. Nevertheless, our minimum energy estimate is far enough beyond the energy of recorded flares that we are motivated to consider whether the event could have been of cometary origin.

  A superflare of energy $10^{34}$ erg could be produced by a solar encounter of a comet (or asteroid) of mass $\gtrsim  10^{19}$ g, which is nearly the mass of C/Hale-Bopp (Weissman, 2007).
  It has been estimated that comet Hale-Bopp has a 0.15 chance of eventually colliding with the sun within several hundred orbits (Bailey, Chanbers, \& Han, 1992, Bailey et al, 1996). Given  that its most recent apparition was less than 2 decades ago, let us assume that a comet this large appears about  twice per century. Multiplying by 0.15, and assuming each one reappears $\sim 300$ times before  colliding with the sun we roughly estimate, on the basis of this very small number statistic, a minimum solar collision rate of $\gtrsim 10^{-5}$ yr$^{-1}$. 

      The above estimate is a lower limit based on a single, long period comet. Of the $\sim 45,000$  Centaurs estimated to exist, whose orbital stability times are of order $10^6$ to $10^7$ yr., many are thought to scatter off Jupiter or outer solar system bodies into the inner solar system, where some meet their demise by crashing into the sun (Horner, Evans \& Bailey, 2004). So a new Centaur enters the inner solar system every  $\sim 10^2$ years or so. If $\sim 0.06$ of them collide with the sun (Levison \& Duncan, 1994), it would be plausible to  expect such an event every $\sim 1.5$ millenia, comparable to the time scale  spanned by the various tree-ring and ice core data sets.

      The Kepler telescope data set reported superflares from 365 stars that displayed brightness variations ($\ge 10^{-3}$) indicative of a magnetic origin (Maehara et al, 2012). In addition, there were 9 superflares from stars, out of a sample of $4.5\cdot 10^4$ stars that  showed otherwise steady emission to within instrumental error (Maehara, private communication). The 
      upper limit on the event rate for stars without periodic luminosity fluctuations greater than $10^{-3}$ is thus about $3\cdot 10^{-4}$ yr$^{-1}$ for all stars, and the question is still open as to what fraction have planetary systems capable of driving  comets into their host stars. We conclude that the hypothesis of a C/Hale-Bopp -size comet hitting the sun every several millenia is consistent with present observations.

\section{Cometary Bow Shocks}
We have also considered a scenario in which the comet converts solar wind energy to energetic particle energy at the bow shock made by its coma. Because the radial ${\bf B}$ field decreases with distance D from the sun as $B(D)\sim 10^{-5}b_5[1\rm A.U./D]^2$G, $b_5\sim 1$, while the radius $R_s$ of the bow shock could increase at most as D, it is clear that $E_o$ decreases with D, and can never be more than $\eta eBDu_{sw}/c\sim 60[1\rm A.U./D]\rm MeV$.  For $E_o$ to be above 300 MeV, D must be at most 0.2 A.U., and the amount of time that the comet would spend within this distance can easily be shown to be about $\delta t = 4\cdot 10^5$s.   
   Assuming the freely streaming energetic particles and the solar wind suffer the same inverse square dilution in getting from the  sunward comet to Earth, it suffices to consider the kinetic energy flux of the solar wind\footnote{with a modest correction for the comet's motion, which adds to $u_s$} (sw) at Earth, $\rho_{sw} u_{s}^3/2$, times $\delta t$. This sets a maximum energy fluence in energetic particles at Earth of  $\rho_{sw} u_{s}^3 \delta t/2 = 7.2 \cdot 10^7  (n_{sw}/{5\rm cm^3})( u_{s}/4\cdot 10^7 \rm cm s^{-1})^3  m_p c^2/\rm cm^2$ and a maximum $^{14}$C production of $\lesssim 10^7$cm$^{-2}$.
  This upper limit falls short of the inferred $^{14}$C production by more than an order  of magnitude. A train of N large comet fragments ($N \gtrsim 10^2$), all from a single progenitor that fragmented,  could each play out the scenario N times within a year, and thus enhance the above estimate.  Moreover, the pitch angle of escaped particles arriving from a sunward source could be biased toward the parallel direction, thus decreasing the amount of mirroring at the poles relative to Galactic cosmic rays, and increasing the fraction of precipitating particles. Nevertheless,
 we believe the original scenario - a C/Hale-Bopp-size comet crashing into the sun -  to be the more conservative, plausible scenario.

       \section{Further Discussion}

       If the most likely explanation of the C.E. 775 event is a superflare at the sun's surface - as opposed to  an event within the solar wind {\it per se} - then the lack of any record of devastation of any sort at that time is reassuring, though not in regard to  safety of power grids and satellites. On the other hand, it implies a huge blast and energetic particle flux, delivered impulsively, at Earth. If it were to repeat in the modern era, it could have devastating consequences for power grids and satellite electronics.

       Most sun-grazing comets are of low mass, $M\lesssim 10^{12}$ g,  and carry  $\lesssim 2\cdot 10^{27}$ erg of kinetic energy into the sun. This is probably not enough to be detectable in high energy particles or their secondaries. Comet Lovejoy, the one known sun-grazer that was large enough to survive the perihelion of a close solar encounter, probably had a surface area of about $10^{10} \, \rm cm^2$ and deposited $ 10^{13}$ g to $ 10^{14}$ g (Sekanina and Chodas, 2012) and $2\cdot 10^{28}$ to $2 \cdot 10^{29}$ ergs during its passage through the corona. Assuming a coma size of $10^9 R_9$ cm, a magnetic field of $B_0$ G, and a shock velocity of $10^{-2.5} \beta_{-2.5} c$,   the value of the cutoff energy $E_o$ is $E_o =  300 R_9 B_0\beta_{-2.5} $ MeV.  It is therefore conceivable that high energy ions, and/or secondary neutrons and gamma rays were produced  at detectable levels during the passage. Results from the IMPACT mission of STEREO are therefore awaited at the time of this writing.  It is possible that at some time in the near future a sun-grazing comet will produce an energetic particle event that will teach us a great deal.
We have also noted that a more extensive data set from the Kepler Observatory could reveal non-magnetic superflares on solar-type stars.

Finally, the $^{10}$Be enhancement during the same event sets an independent constraint on the spectrum of energetic particles that caused the CE 775 event (Miyake et al. 2012). Though beyond the scope of this paper, a careful analysis of the $^{10}$Be data during that time  could test our rough estimate for the spectral cut-off parameter $E_o\gtrsim 1/3$ GeV, and the attendant value of the $^{14}$C production efficiency $\epsilon$.

      I thank R. Kumar for his kind help in the numerical integration, Prof. H. Maehara for sharing unpublished Kepler data,   Prof. N. Shaviv, Prof. N. Brosch, Prof. D. Prialnik, Dr. M. Eichler, for helpful conversations, and the support of the Israel-U.S. Binational Science Foundation, the Israeli Science Foundation, and the Joan and Robert Arnow Chair of Theoretical Astrophysics.

      \bigskip
      \bigskip

\centerline{\bf REFERENCES}

\noindent Bailey, M. E., Chambers, J. E. \& Hahn, G., 1992, Astron. and Astroph., 257, 315

\noindent Bailey, M. E., Emel'yanenko, V.V., Hahn, G.,  Harris, N.W., Hughes, K.A.,  Muinonen, K., \& Scotti, J.V., 1996, 
Monthly Notices of the Royal Astronomical Society, 281, 916

\noindent Brown, J.C., Potts, H.E.,  Porter, L.J.  \&  le Chat, G., 2011, Astronomy and Astrophysics, 535A, 71B

\noindent Chang, S.W., Scudder, J.D., Kudela, K., Spence, H. E., Fennell, J. F., Lepping, R. P., Lin, R. P., Russell, C. T., 2001, JGR, 106,  19101C


\noindent  Decker, R.B., Krimigis, S.M., Roelof, E.C., Hill M.E., Armstrong. T. P., Gloeckler, G., Hamilton, D.C.,  Lanzerotti, L.J., 2008,  Nature, 454, 67

\noindent Eichler, D. 1979, ApJ, 229, 4

\noindent Eichler, D. 1981a, ApJ. 244, 711

\noindent Eichler, D. 1981b, ApJ. 247, 189

\noindent Eichler,  D. 1985, ApJ. 294, 40

\noindent Ellison, D.C. and Eichler, D., 1985, Phys. Rev. Lett., 55, 273

\noindent Ellison, D.C. \& Mobius, E.,    1987, ApJ,  318, 474

\noindent Horner, J. Evans, N.W. \& Bailey, M.E., 2004, M.N.R.A.S., 354, 798

\noindent Kovaltsov, G.A., Mishev, A. \& Usoskin, I.G.,  2012, Earth Planet. Sci. Lett.  337, 114

\noindent Levison, H.F., \& Duncan, M.J., 1994, Icarus, 108, 18


\noindent Maehara, H., et al.  2012, Nature, 485, 478

 \noindent Miyake,F. Nagaya, K.  Masuda, K. \& and Nakamura, T. , Nature, 2012, 486, 240




 \noindent Schrijver, C.J.,  Brown, J.C., Battams, K., Saint-Hilaire,P., Liu,W.,
Hudson, H., \& Pesnell, W.D., 2012, Science, 335, 324

 \noindent Sekanina, Z. \& Chodas, P.W.,  2012,  ApJ., 757, 127

 \noindent Weissman, P. R., 2007, Proceedings of the International Astronomical Union 2 (S236), 441 (Cambridge University Press, {\it ed}: Milani, A., Valsecchi, G.B. \& Vokrouhlicky, D.)

    \end{document}